\documentclass[journal,a4paper,twoside]{IEEEtran}

\usepackage[]{amsmath}
\usepackage{mathrsfs}
\usepackage{dcolumn}
\usepackage{bm}

\usepackage{graphicx}
 \usepackage[font=footnotesize, caption=false]{subfig}
\usepackage{epsf}
\usepackage{psfrag}
\usepackage{dcolumn}
\usepackage{flushend}

\makeatletter
\long\def\@makecaption#1#2{\ifx\@captype\@IEEEtablestring%
\footnotesize\begin{center}{\normalfont\footnotesize #1}\\
{\normalfont\footnotesize\scshape #2}\end{center}%
\@IEEEtablecaptionsepspace
\else
\@IEEEfigurecaptionsepspace
\setbox\@tempboxa\hbox{\normalfont\footnotesize {#1.}~~ #2}%
\ifdim \wd\@tempboxa >\hsize%
\setbox\@tempboxa\hbox{\normalfont\footnotesize {#1.}~~ }%
\parbox[t]{\hsize}{\normalfont\footnotesize \noindent\unhbox\@tempboxa#2}%
\else
\hbox to\hsize{\normalfont\footnotesize\hfil\box\@tempboxa\hfil}\fi\fi}
\makeatother

\makeatletter
\def\ps@headings{%
\def\@oddhead{\mbox{}\scriptsize\rightmark \hfil }%
\def\@evenhead{\scriptsize \hfil \leftmark\mbox{}}%
\def\@oddfoot{}%
\def\@evenfoot{}}
\makeatother

\makeatletter
\def\ps@IEEEtitlepagestyle{%
\def\@oddhead{\mbox{}\scriptsize\rightmark \hfil }%
\def\@evenhead{\scriptsize \hfil \leftmark\mbox{}}%
\def\@oddfoot{}%
\def\@evenfoot{}}
\makeatother

\begin{document}

\title{Non-Equilibrium Superconductivity in Kinetic Inductance Detectors for THz Photon Sensing}

%

\author{\IEEEauthorblockN{D. J. Goldie\IEEEauthorrefmark{1} and S. Withington\IEEEauthorrefmark{1}}\\
\IEEEauthorblockA{\IEEEauthorrefmark{1}Detector and Optical Physics Group\\
Cavendish Laboratory\\ University of Cambridge\\
JJ Thomson Av.\\ Cambridge\\CB3 0HE, UK \\
Email: d.j.goldie@mrao.cam.ac.uk}\\
}

\markboth{24th International Symposium on Space Terahertz Technology, Groningen, 8-10 April, 2013.}%
{24th International Symposium on Space Terahertz Technology, Groningen, 8-10 April, 2013.}

\maketitle

\begin{abstract}
Low temperature Kinetic Inductance Detectors (KIDs) are attractive candidates for producing quantum-sensitive,
 arrayable  sensors for astrophysical and other precision measurement applications.
The readout uses a low frequency probe signal with quanta of energy well-below the threshold for pair-breaking in the superconductor.
  We have calculated the detailed non-equilibrium quasiparticle and phonon energy spectra
generated by the  probe signal of the KID when
operating well-below its superconducting transition temperature $T_c$
within the framework of the coupled kinetic equations described by Chang and Scalapino.\cite{Chang_and_Scalapino_ltp}
At the lowest bath temperature studied $T_b/T_c=0.1$ the quasiparticle distributions can be driven far from equilibrium.
In addition to the low frequency probe signal we have incorporated a high frequency ($\sim 1\,\,{\rm THz}$) source signal well-above the
pair-breaking threshold of the superconductor.
Calculations of source signal detection efficiency are discussed.
\end{abstract}


\section{Introduction}

Kinetic inductance detectors  (KIDs)  operating at low reduced temperatures $T/T_c\simeq 0.1$, where
 $T$ is the temperature and $T_c$ is the superconducting
transition temperature, are used not only
as ultra-sensitive   detectors of incident power or individual quanta
for applications in sub-millimeter, millimeter, optical, X- and $\gamma$-ray astrophysics,\cite{Jonas_nature, Jonas_review, George_kid, Monfardini, Baselmans_review, deVisser_apl_2012}
but also
as elements of Qubits for
quantum computing.\cite{Dicarlo_nature, Hofheinz_nature, Schoelkopf_nature}
As a detector the superconductor is formed as a resonator and  changes in its complex conductance can be monitored by measuring the complex
transmission $S_{21}$ of a probe signal.
We recently described a detailed microscopic calculation of the spectrum of the non-equilibrium quasiparticles and phonons in
a KID operating at $T/T_c=0.1$.\cite{Goldie_SuST_2013}
Prior to that work and despite the technological importance, a detailed microscopic analysis  of the effect on the distribution functions of the 
quasiparticles and phonons at temperatures  $T\sim 0.1 T_c$ due to the interaction of a {\it flux} of microwave photons of frequency $\nu_p\ll 2\Delta(T)/h$, where $2\Delta(T)$ is the temperature-dependent  superconducting energy gap
and $h$ is Planck's constant, seemed to be lacking. By contrast the regime $\nu_p\sim \Delta(T)/h$ with $T\sim T_c$, when gap enhancement effect are predicted and observed, has been extensively studied.
The quasiparticles and phonons of a low temperature superconductor form coupled subsystems.
Energy relaxation processes of non-equilibrium quasiparticles  comprise
scattering  with absorption or emission of phonons, and scattering involving
Cooper pairs with
generation or  loss of two  quasiparticles  and absorption or
emission of phonons of energy $\Omega\ge 2\Delta$ respectively. Energy escapes from the superconductor as phonons enter the substrate.
 The coupled kinetic equations that describe these interacting subsystems were derived by Bardeen, Rickayzen and Tewordt\cite{Bardeen_Rickayzen}
 and
discussed in detail by Chang and Scalapino.\cite{Chang_and_Scalapino_ltp, Chang_and_Scalapino_prb}
The  coupled kinetic equations have been used  to investigate the
effect of high energy photon interactions ($h\nu_{sig}/\Delta \sim 3\times 10^7$) at $T/T_c\sim 0.1$.\cite{Zehnder_model,Ishibashi}

 In Ref.~\cite{Chang_and_Scalapino_ltp}
full non-linear
solutions were obtained which is the approach we have adopted. Crucially however in that earlier work solutions were obtained close to $T_c$ where
microwave drive can lead to gap-enhancement effects. In the present programme our interest lies in the behavior at  low effective temperatures,
where changes in the quasiparticle density have most effect on the KID.
The KID is readout with a  microwave probe signal of energy $h\nu_p\ll 2\Delta$ where $\nu_p$ is the probe frequency close to the
resonant frequency of the KID.
Our fundamental observation is that the readout is dissipative, but that there are very few {\it thermal}
quasiparticles present at  $T_b$
which can  interact with the probe. Our solutions of the coupled kinetic equations showed that the KID can be driven far from equilibrium for typical
experimental probe powers.\cite{Goldie_SuST_2013}
 Here we begin to explore the effect of adding a signal power comprising photons of energy $hv_{sig}\ge2\Delta$ so that the signal
breaks Cooper pairs in addition to the probe signal for which multiple photon absorption breaks pairs.

An important consideration in the  design of a KID for THz photons is the fraction of incident signal power (or indeed energy for single quantum detection)
that is coupled to the quasiparticles. The detection geometry we consider would allow the signal to interact directly in the
superconductor and so that the signal breaks Cooper pairs.
Pair breaking creates excess (primary) quasiparticles, and these quasiparticles scatter
to lower energies emitting phonons on a timescale that is
on average shorter than the effective population recombination time.
 These phonons
will be lost from the KID if $\Omega<2\Delta$ but may break additional Cooper pairs if $\Omega\ge2\Delta$. Pair breaking
 increases
the total number of quasiparticles created by the initial photon interaction and hence the signal that is detected.
Some fraction of the pair-breaking phonons will still be lost from a  KID of finite thickness a process
which reduces the overall detection efficiency.
The probability of pair-breaking is determined by the phonon pair breaking time $\tau_{pb}$ and the
phonon loss time from the film $\tau_{loss}$. At low temperature and low phonon energies $\tau_{pb}=\tau_0^\phi$ where
$\tau_0^\phi$ is the characteristic phonon lifetime.\cite{Kaplan}

Kurakado\cite{Kurakado} used the equilibrium lifetimes described by Kaplan {\it et al.}\cite{Kaplan} to
 describe the interaction of a single excess phonon or  quasiparticle in a {\it bulk} superconductor at $T/T_c=0$
finding that the average energy required to create a quasiparticle was $\epsilon=1.68\Delta$, or equivalently an efficiency $\eta=0.59$, where the excess quasiparticles are  assumed to have $E=\Delta$.\cite{Kurakado}
Obviously, because an infinite superconductor is modeled,  phonon loss is ignored unless $\Omega<2\Delta$, and likewise recombination. The effect of a thermal (or even driven) population  is likewise  ignored since $T/T_c=0$.
Zehnder investigated the interaction of photons of energy ($h\nu_{sig} \sim 3\times10^7 \Delta$)  in a number of thin film superconductors at
$T/T_c=0.1$ including quasiparticle diffusion and phonon loss
although did not extend the modeling to low incident photon energies of interest here.

To our knowledge no solutions of the full coupled equations  exist of the  efficiency with which
monochromatic photons with $h\nu_{sig} \sim 2-30\Delta$
create quasiparticles in a thin-film superconductor  {\it including} $2\Delta$-phonon loss with a probe signal which itself breaks
pairs through  multiple photon processes.
\section{Non-equilibrium KIDs}
\label{sec:TES_thermometry}
The coupled non-linear equations described by Chang and Scalapino were solved using
 Newton-Raphson iteration. Details of the scheme, the representation of the
 quasiparticle and phonon distributions, and the convergence criteria are given in
Ref.~\cite{Goldie_SuST_2013}.
In this way non-equilibrium quasiparticle and phonon energy distributions
 $f(E)$ and $n(\Omega)$ can be calculated. $E$ and $\Omega$ are the quasiparticle and
phonon energies respectively.
An approach to find the drive term of the quasiparticles
$I_{probe}$  associated with the probe power was also described, based
on the assumption that the absorbed probe power per unit volume $P_{probe}$ can be measured experimentally.
Here we adopt a similar approach to calculate the effect of an additional signal power per unit volume
$P_{sig}$.
\subsection{Including a pair-breaking signal }
\label{subsec:Signal}
The effect of a signal with photons of energy $E = h\nu_{sig}$ and absorbed power per unit volume
$P_{sig}$
can be included in a similar way to the probe signal.
\begin{figure}[!t]
 \begin{center}
   \begin{tabular}{c}
   \psfrag{Xaxis} [] [] { $E /\Delta  $ }
   \psfrag{Yaxis} [] [] { $ K_{sig} \rho(E,\Delta)  $  }
   \includegraphics[height=8.5cm, angle=-90]{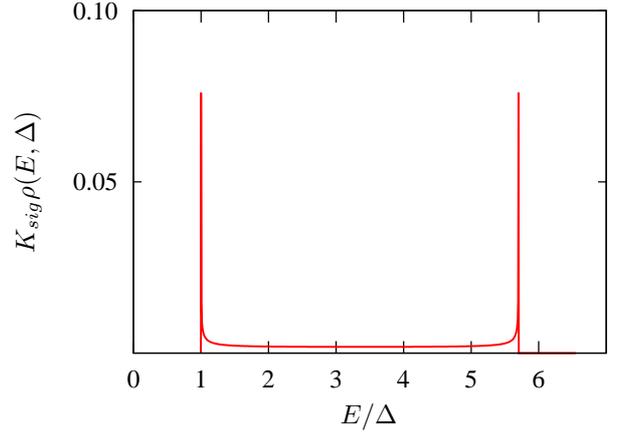}
   \end{tabular}
  \end{center}
   \caption[Fig1]
   { \label{fig:K_qp}
 (Color online) The (number) drive term $ K_{sig} \rho(E,\Delta)  $  for $h\nu_{sig}=6.67\Delta$. }
   \end{figure}
The signal  contributes  an additional drive term to Eq.~[2] of Ref.~\cite{Goldie_SuST_2013}
for the quasiparticle distribution function
\begin{equation}
 \delta f(E)/\delta t \vert_{sig} = I_{sig} ,
\label{Eq:Photon_drive}
\end{equation}
where $I_{sig}=B_{sig}K_{sig}$,
\begin{equation}
\begin{split}
K_{sig}(E,\nu_{sig}) = & \\
			2    \Bigg[ & \rho(E+h\nu_{sig},\Delta)
\left[ 1+ \frac {\Delta^2}{E\left(E+h\nu_{sig}\right)} \right]\\
& \left[f\left(E+h\nu_{sig}\right)-f\left(E\right)\right] \\
  - & \rho(E-h\nu_{sig},\Delta)
 \left[1 +\frac {\Delta^2}{E\left(E-h\nu_{sig}\right)} \right] \\
& \left[f \left(E\right) - f\left(E-h\nu_{sig}\right)\right] \\
+ & \rho(h\nu_{sig} -E,\Delta)\left[ 1- \frac {\Delta^2}{E\left(h\nu_{sig}- E \right)} \right] \\
 & \left[1 - f\left(E\right) -f\left(h\nu_{sig}-E\right) \right]   \Bigg] ,
\label{Eq:Drive_1}
\end{split}
\end{equation}
and the prefactor $B_{sig}$ normalizes the signal power absorption so that
\begin{equation}
B_{sig}(\nu_{sig})= \frac{P_{sig}}{  4 N_0 \int_\Delta^\infty E \rho(E) K_{sig}(E,\nu_{sig}) dE         } .
\label{Eq:B_sig}
\end{equation}
In the results discussed later we use as an example  a signal with photon energy $h\nu_{sig}=6.67\Delta$ corresponding to an absorbed frequency
$\nu_{sig}=290\,\,{\rm GHz}$ in Al.
Eq.~\ref{Eq:Drive_1} differs from that to describe the probe power in having a third term. This term represents  pair-breaking
and occurs provided $h\nu_{sig}\ge 2\Delta$. At low temperatures (and low probe powers) this term is the dominant contribution to $I_{sig}$.
Figure~\ref{fig:K_qp}
 shows $K_{qp}$  multiplied by $\rho(E,\Delta)$, thus showing the contribution to the number change,
for a pair-breaking signal at low temperatures normalized so that each absorbed photon produces two quasiparticles.
The double peak arises because the quasiparticle number
generated by pair breaking involves the {\it product}
of final state densities $\rho(E_{sig}-E^\prime,\Delta)\rho(E^\prime,\Delta)$. The density of states is peaked at $E=\Delta$ and the product is symmetric
with respect to the final state energies.
\subsection{KID model parameters }
\label{subsec:Model}
We have used the same parameters to describe the KID given in Ref.~\cite{Goldie_SuST_2013}
which are appropriate for Al.
We used
 $\Delta(0)=180\,\,{\rm \mu eV}$, $T_c=1.17\,\,{\rm K}$  and we set $T_b/T_c=0.1$.
The single spin density of states was $N(0)=1.74\times 10^{4} \,{\rm \mu eV^{-1} \mu m^{-3}}$, characteristic quasiparticle time
$\tau_0=438\,\,{\rm ns}$ and the characteristic phonon
 (Debye model) lifetime $\tau_0^\phi=0.26\,\,{\rm ns}$.\cite{ Goldie_SuST_2013,Kaplan}
In all calculations we assume that the phonon loss can be characterized by a single energy independent time and we assume
$\tau_{loss}/\tau_0^\phi = 1$ which we
estimate would be appropriate for a $70\,\,{\rm nm}$ Al film on Si.\cite{Kaplan_loss}
We assume a probe photon energy $h\nu_{probe}=16\,\,{\rm \mu eV}$, ($\nu_{probe}= 3.88\,\,{\rm GHz}$).
\section{Results}
\label{sec:Results}
Here we show results of the numerical modeling.
\begin{figure}[!t]
 \begin{center}
   \begin{tabular}{c}
    \psfrag{Xaxis} [] [] { $E /\Delta  $ }
    \psfrag{Yaxis} [] [] { $ f(E)  $  }
   \psfrag{aa} [r] [r] [0.75] { $h\nu_{sig}  $ }
   \includegraphics[height=8.5cm, angle=-90]{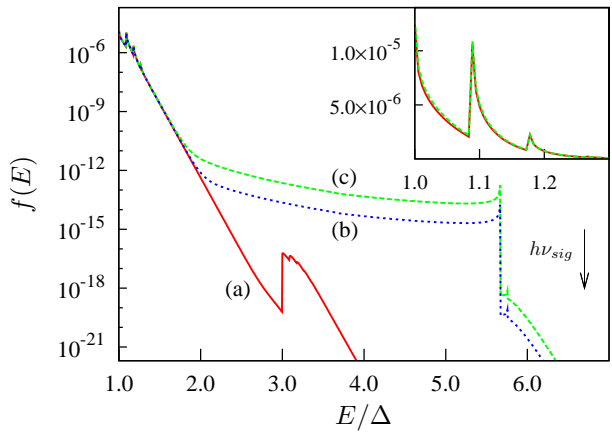}
   \end{tabular}
  \end{center}
  \caption[Fig1]
   { \label{fig:f_with_signal}
 (Color online) Semi-log plot showing the effect of   $P_{probe}=0.5\,\,{\rm aW/\mu m^3}$ on the quasiparticle distributions;
  (a) probe power only (full line red line),  (b) $P_{sig}/P_{probe}=0.01  $ (dashed green line) and (c) $P_{sig}/P_{probe}=0.001  $ (dashed blue line) both
  with the same probe power. The  drive photon energy $h\nu_{sig}=6.67\Delta$ and $\tau_{loss}/\tau_0^\phi = 1$.
The inset shows the same $P_{probe}=0.5\,\,{\rm aW/\mu m^3}$ and  $P_{sig}/P_{probe}=0.01  $
at low energies with a linear ordinate to emphasize the changes. }
   \end{figure}
 Fig.~\ref{fig:f_with_signal} shows the calculated non-equilibrium quasiparticle distributions for a probe power of
$P_{probe}=0.5\,\,{\rm aW/\mu m^3}$ having a probe photon energy $h\nu_{p}=16\,\,{\rm \mu eV}$, as the solid curve
and also the additional effect of a pair-breaking signal of power $P_{sig}/P_{probe}=0.01  $ (dashed green curve)
 and $P_{sig}/P_{probe}=0.001$ (dashed blue curve). The inset shows a low energy detail of the distributions created by
the probe itself and the signal of  $P_{sig}/P_{probe}=0.01  $.
The main figure shows a number of effects. For the probe signal alone  at low energies $E/ \Delta \sim 1 $ we see the multiple
peaked structure corresponding to absorption of the probe signal by the large density of quasiparticles near the gap.
At energies $E/ \Delta \sim 3$ we see a step in the distribution corresponding to reabsorption of non-equilibrium pair-breaking phonons by the
quasiparticles. This structure also exhibits peaks associated with multiple photon absorption from the probe.
The distribution functions calculated with an additional pair-breaking signal have similar structure at low energies but show a
step in the distribution at $E=h\nu_{sig}-\Delta$.
\begin{figure}[!t]
 \begin{center}
  \begin{tabular}{c}
   \psfrag{Xaxis} [] [] { $\Omega/ \Delta  $ }
	\psfrag{Yaxis} [] [] { ${ \delta P(\Omega)_{\phi-b} }\,\, ( \rm{ kHz/  \mu m^3} )  $ }
   \psfrag{A} [r] [r] [0.85] { $P_{sig}=0.1P_{probe} $ }
   \psfrag{B} [r] [r] [0.85] { $P_{sig}=0.01P_{probe}  $ }
   \psfrag{ab} [r] [r] [0.75] { $\Omega= h\nu_{sig} -2 \Delta $ }
   \includegraphics[height=8.5cm, angle=-90]{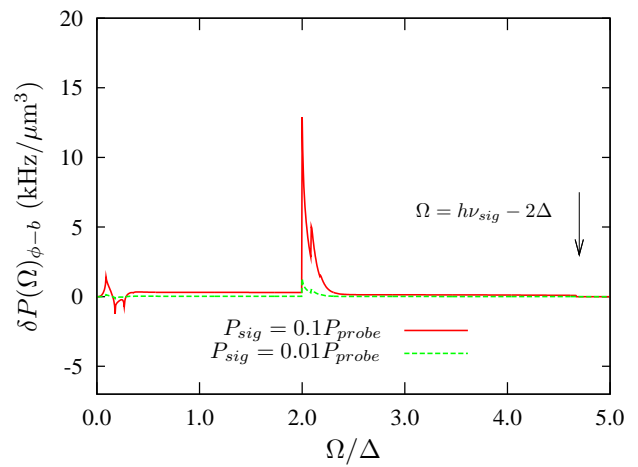}
   \end{tabular}
  \end{center}
   \caption[Fig3]
   { \label{fig:dP_Phonon_sig}
 (Color online) The change in the phonon power flow to the bath $\delta P(\Omega)_{\phi-b}$   for
   $P_{sig}=0.1P_{probe} $ and $P_{sig}=0.01P_{probe} $ for $P_{probe}=0.5\,\,{\rm aW/\mu m^3}$.
   The  drive photon energy $h\nu_{sig}=6.67\Delta$ and $\tau_{loss}/\tau_0^\phi = 1$.
 }
   \end{figure}
This peak  is expected due to the high density of available states at
$E=\Delta$, and the curvature of the distribution below this peak arises from the energy dependence of the
quasiparticle scattering and recombination rates. The photon peak itself also has a smaller ``satellite" peak at
$E=h\nu_{sig}-\Delta + h \nu_p $. This arises from absorption of the probe  by the quasiparticles created by the signal photons.
The inset shows the detail at low energies. It is (just) possible to observe that the distribution with signal is enhanced over that of the probe alone.

Fig.~\ref{fig:dP_Phonon_sig} shows the {\it change}  in the phonon power flow to the heat bath after subtraction of  that without the signal for two signal powers.
We have plotted
$\delta P(\Omega)_{\phi-b}= P(\Omega)^{sig}_{\phi-b}- P(\Omega)^{probe}_{\phi-b}$, where
$P(\Omega)^{sig}_{\phi-b}$ is the contribution to the phonon-bath power flow with signal and probe, and $P(\Omega)^{probe}_{\phi-b}$
that for the probe alone.
The power flow contributions are most easily seen in the plot corresponding to the higher signal power. At low phonon
energies $\Omega/ \Delta <0.15 $ corresponding to the first probe photon peak there is an increase in the power flow to the bath. At energies
$0.15< \Omega/\Delta < 0.29 $ the power is reduced. The first effect is expected as the signal itself has a sharply peaked structure
near the gap. The reduction at slightly higher energies is at first sight more surprising
 but arises from the blocking of final states for the  scattering of the higher energy
probe-generated quasiparticle peaks towards the gap.

At higher phonon energies there is a significant change in the contribution to the power flow from
recombination phonons $\Omega/\Delta \ge 2$ as would be expected for a pair-breaking detection.
The energy spectrum also shows a broad low background contribution at all phonon energies
$\Omega/ \Delta\le \left(h\nu_{sig}-2\Delta \right) /\Delta$
corresponding to phonons generated by quasiparticle scattering to final state energies $E\sim \Delta$.

\section{Photon Detection Efficiency}
\label{sec:Detection_Efficiency}
Here we  quantify the overall quasiparticle creation efficiency using a simple rate equation approach. We have assumed that an incident monochromatic signal of
power per unit volume $P_{sig}$ is absorbed by the quasiparticles.
\begin{figure}[!t]
 \begin{center}
   \begin{tabular}{c}
   \psfrag{Xaxis} [] [] { $E /\Delta  $ }
   \psfrag{Yaxis} [] [] { $ K_{sig} \rho(E,\Delta)  $  }
   \includegraphics[height=8.5cm, angle=-90]{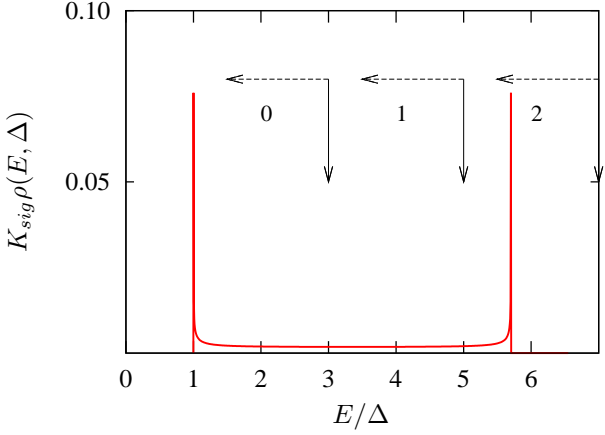}
   \end{tabular}
  \end{center}
   \caption[Fig1]
   { \label{fig:Eff_model}
 (Color online) The (number) drive term $ K_{sig} \rho(E,\Delta)  $  for $h\nu_{sig}=6.67\Delta$. }
   \end{figure}
Only a fraction $\eta_{sig}$ of the absorbed signal
 supports the excess quasiparticle density $N_{ex}$ 
 because some fraction of the phonons generated in the down-conversion
are lost into the substrate.
During the down-conversion the total number of quasiparticles can however be
 increased by reabsorption of those phonons which  break pairs.
 If on average each absorbed signal photon generates $m_{sig}$ additional quasiparticles then  the sum of the rates of generation by photons and loss by phonon processes is given by
\begin{equation}
\frac{\delta N^{ex}_{qp}}{\delta t}=m_{sig} \frac{\delta N_{sig}}{\delta t} + \left.{\frac{\delta N^{ex}_{qp} }{\delta t}}\right \vert_\phi.
\label{Eq:simple_rate}
\end{equation}
  In steady state
  $\delta N^{ex}_{qp}/\delta t=0$ and we use  $N^{ex}_{qp} = 4 N_0\int_\Delta^\infty \rho(E) (f_{sig}-f_{p}) dE$.
The rate of  photon absorption from the signal is $\delta N_{sig}/\delta t =P_{sig}/h \nu_{sig}$.
 The loss rate due to phonons is
$ \delta N^{ex}_{qp}/\delta t\vert_\phi = -N^{ex}_{qp}/ \tau_r^{eff} $ so we have
\begin{equation}
m_{sig} =  \frac{ h \nu_{sig} 4 N_0 \int_\Delta^\infty \rho(E) (f_{sig}-f_{p}) dE }{P_{sig} \tau_r^{eff}},
\label{Eq:m_sig}
\end{equation}
where $\tau_r^{eff}=\tau_r^{sig}/2\left(1+\tau_l/\tau_{pb}\right)$ is the effective recombination time for the driven population
with  the signal.
We find that $m_{sig} = 3.29$ for $h \nu_{sig}= 6.67\Delta$ giving a number detection
efficiency $\eta_n= 3.29\Delta/6.67\Delta=0.49$ if we assume that all of the signal-generated excess  quasiparticles
in static non-equilibrium  have $E=\Delta$.
In a similar way we can calculate the power detection
 efficiency
\begin{equation}
\eta_{sig}=\frac{ 4 N_0\int_\Delta^\infty E \rho(E) (f_{sig}-f_{p}) dE} { P_{sig}\tau_r^{eff}  }
\label{Eq:eta_sig}
\end{equation}
giving $\eta_{sig}=0.51$ and here we have taken  account of the energy distribution of the excess quasiparticles.
\section{Discussion and Conclusions}
We have presented preliminary calculations of the detailed energy spectra of the non-equilibrium quasiparticles and phonons of
a representative and technologically interesting low temperature superconductor (here Al)
generated by a low power pair-breaking signal of
frequency $\nu_{sig}=290\,\,{\rm GHz}$. Considering the energy relaxation processes
within the superconductor the calculation is fully representative of
photon absorption up to $h\nu_{sig}=\Omega_D$ where $\Omega_D$ is the Debye energy,
and for reference $\Omega_D\sim 8 \,\,{\rm THz}$.
The model includes a higher power  probe of frequency
 $\nu_{probe}\sim 4\,\,{\rm GHz}$ chosen to be typical of the powers and frequencies used in KID readout.
 The detailed spectra  show the effects of interaction between the probe and the signal
 showing structure for example at $E=h(\nu_{sig}+\nu_{probe})$.
In future work we will extend the model to investigate the detection linearity of a resonator with the
driven distributions. It will also be possible for example to  calculate the behaviour
 of the resonator used as a mixer.

We calculated the effective population quasiparticle lifetime
for the driven distribution and used a simple rate equation approach to find the
static driven number of quasiparticles generated by the high frequency signal.
In this way a number detection efficiency $\eta_{sig}\sim 0.5$ was found
for a signal of frequency $\nu_{sig}=290\,\,{\rm GHz}$
assuming a phonon loss time $\tau_{loss}/\tau_0^\phi = 1$.
This efficiency may seem at first sight low.
Figure~\ref{fig:Eff_model}
presents a naive model to understand the energy down-conversion and appreciate the
calculated detection efficiency. The curve reproduces the quasiparticle number spectrum generated by a single
interacting signal photon shown in Fig.~\ref{fig:K_qp}.
The vertical arrows at $3$, $5$ and $7\Delta$ are intended to break the number spectrum into regions
labeled 0, 1 and 2 respectively. In the simplest approximation we would assume that
quasiparticles relax by scattering i.e. ignoring recombination.
 The group of quasiparticles in region 0 have energies below $3\Delta$. On scattering to lower energies,
predominantly $E=\Delta$,
they emit phonons of energy $\Omega<2\Delta$ which are lost from the film. Scattering of this group
 does not change the static number density
merely the spectrum. Quasiparticles in region 1 have energies  $3\Delta<E<5\Delta$. On scattering to lower energies,
these emit secondary phonons of energy $2\Delta<\Omega<4\Delta$.  These may break pairs and the probability of pair breaking over
all possible phonon processes is $p=\tau_{loss}/(\tau_{loss}+\tau_{pb}) $. Ignoring the energy dependence of $\tau_{pb}$ then
$p=0.5$, (in this approximation $\tau_{pb}=\tau_0^\phi$\cite{Kaplan}), so that a fraction of the power
 generating  these quasiparticles would
be lost from the film during the down-conversion. A similar discussion would apply to region 3 but now the secondary phonons
with $4\Delta<\Omega<6\Delta$
create secondary quasiparticles with energies $3\Delta<E<5\Delta$ and probability $p$. These in turn scatter and
 create tertiary pair-breaking phonons
 of which a fraction $p$ create additional pairs. The overall probability of this process is reduced ($p^\prime=p^2$). We have not as yet set up a detailed
model of this process within this framework. This would need to include not only the energy dependence of $\tau_{pb}(\Omega)$ but also the
spectrum of phonons generated as the primary quasiparticle spectrum relaxes and the resultant spectrum of the secondary quasiparticles.
To an extent  this extended (naive) calculation should begin to
approximate the detail contained in the full non-equilibrium solutions already described.
 Even so, using even the very simplest approach, approximating the spectrum of Fig.~\ref{fig:Eff_model} by
$\delta$-functions at  $E=\Delta$ and $E=h\nu_{sig}-\Delta$, we estimate  $\eta\sim 0.52$ in excellent agreement with the full
non-equilibrium calculation and we would expect recombination to reduce this estimate. For higher incident photon energies we would expect this efficiency to be further reduced.

We believe that the model we have described and in particular the detection efficiency of thin-film superconductors
 in the ${\rm THz}$
regime has important consequences. If the quasiparticle creation efficiency is as we have described, the
achievable sensitivity, or equivalently noise equivalent power of KIDs used for this application in the geometry considered may be compromised, certainly if earlier published estimates are used which ignore phonon loss from thin films.




\bibliography{TESreferences5}   

\begin{thebibliography}{10}
\providecommand{\url}[1]{#1}
\csname url@samestyle\endcsname
\providecommand{\newblock}{\relax}
\providecommand{\bibinfo}[2]{#2}
\providecommand{\BIBentrySTDinterwordspacing}{\spaceskip=0pt\relax}
\providecommand{\BIBentryALTinterwordstretchfactor}{4}
\providecommand{\BIBentryALTinterwordspacing}{\spaceskip=\fontdimen2\font plus
\BIBentryALTinterwordstretchfactor\fontdimen3\font minus
  \fontdimen4\font\relax}
\providecommand{\BIBforeignlanguage}[2]{{%
\expandafter\ifx\csname l@#1\endcsname\relax
\typeout{** WARNING: IEEEtran.bst: No hyphenation pattern has been}%
\typeout{** loaded for the language `#1'. Using the pattern for}%
\typeout{** the default language instead.}%
\else
\language=\csname l@#1\endcsname
\fi
#2}}
\providecommand{\BIBdecl}{\relax}
\BIBdecl

\bibitem{Chang_and_Scalapino_ltp}
J.~J. Chang and D.~J. Scalapino, ``Nonequilibrium superconductivity,'' \emph{J.
  Low Temp. Phys.}, vol.~31, pp. 1--32, 1978.

\bibitem{Jonas_nature}
P.~K. Day, H.~G. {LeDuc}, B.~A. Mazin, A.~Vayonakis, and J.~Zmuidzinas, ``A
  broadband superconducting detector suitable for use in large arrays,''
  \emph{Nature}, vol. 425, pp. 817--821, 2003.

\bibitem{Jonas_review}
J.~Zmuidzinas, ``{Superconducting Microresonators: Physics and Applications},''
  \emph{{Ann. Rev. Condens. Matter Phys.}}, vol.~{3}, pp. 169--214, {2012}.

\bibitem{George_kid}
G.~Vardulakis, S.~Withington, and D.~J. Goldie, ``{Superconducting kinetic
  inductance detectors for astrophysics},'' \emph{Meas. Sci. Technol.},
  vol.~19, p. 015509, 2008.

\bibitem{Monfardini}
A.~Monfardini, L.~J. Swenson, A.~Bideaud, F.~X. Desert, S.~Doyle, B.~Klein,
  M.~Roesch, C.~Tucker, P.~Ade, M.~Calvo, P.~Camus, C.~Giordano, R.~Guesten,
  C.~Hoffmann, S.~Leclercq, P.~Mauskopf, and K.~F. Schuster, ``{NIKA: A
  millimeter-wave kinetic inductance camera},'' \emph{{Astron. Astrophys.}},
  vol. {521}, p. {A29}, {2010}.

\bibitem{Baselmans_review}
J.~J.~A. Baselmans, ``{Kinetic Inductance Detectors},'' \emph{J. Low Temp.
  Phys.}, vol. 167, pp. 292--304, 2011.

\bibitem{deVisser_apl_2012}
P.~J. {de Visser}, J.~J.~A. Baselmans, S.~J.~C. Yates, P.~Diener, A.~Endo, and
  T.~M. Klapwijk, ``{Microwave-induced excess quasiparticles in superconducting
  resonators measured through correlated conductivity fluctuations},''
  \emph{Appl. Phys. Lett.}, vol. 100, p. 162601, 2012.

\bibitem{Dicarlo_nature}
L.~{DiCarlo}, M.~D. Reed, L.~Sun, B.~R. Johnson, J.~M. Chow, J.~M. Gambetta,
  L.~Frunzio, S.~M. Girvin, M.~H. Devoret, and R.~J. Schoelkopf, ``{Preparation
  and measurement of three-qubit entanglement in a superconducting circuit},''
  \emph{Nature}, vol. 467, pp. 575--578, 2010.

\bibitem{Hofheinz_nature}
M.~Hofheinz, E.~M. Weig, M.~Ansmann, R.~C. Bialczak, E.~Lucero, M.~Neeley,
  A.~D. {O'Connell}, H.~Wang, J.~M. Martinis, and A.~N. Cleland, ``{Generation
  of Fock states in a superconducting quantum circuit},'' \emph{Nature}, vol.
  454, pp. 310--314, 2008.

\bibitem{Schoelkopf_nature}
R.~V. Schoelkopf and S.~M. Girvin, ``{Wiring up quantum systems},''
  \emph{Nature}, vol. 451, pp. 664--669, 2008.

\bibitem{Goldie_SuST_2013}
D.~J. Goldie and S.~Withington, ``{Non-equilibrium superconductivity in
  quantum-sensing superconducting resonators},'' \emph{{Supercond Sci and
  Tech}}, vol.~{26}, p. 015004, {2013}.

\bibitem{Bardeen_Rickayzen}
J.~Bardeen, G.~Rickayzen, and L.~Tewordt, ``{Theory of the thermal conductivity
  of superconductors},'' \emph{Phys. Rev.}, vol. 113, pp. 982--994, 1959.

\bibitem{Chang_and_Scalapino_prb}
J.~J. Chang and D.~J. Scalapino, ``Kinetic-equation approach to
  superconductivity,'' \emph{Phys. Rev. B}, vol.~15, pp. 2651--2670, 1977.

\bibitem{Zehnder_model}
A.~Zehnder, ``{Response of superconducting films to localized energy
  deposition},'' \emph{Phys. Rev. B}, vol.~52, pp. 12\,858--12\,866, 1995.

\bibitem{Ishibashi}
K.~Ishibashi, K.~Takeno, T.~Nagae, and Y.~Matsumoto, ``{Output signal from
  Nb-based tunnel junctions by irradiation of 6 keV X-rays},'' \emph{IEEE
  Trans. Magnetics}, vol.~27, pp. 2661--2664, 1991.

\bibitem{Kaplan}
S.~B. Kaplan, C.~C. Chi, D.~N. Langenberg, J.~J. Chang, S.~Jafarey, and D.~J.
  Scalapino, ``Quasiparticle and phonon lifetimes in superconductors,''
  \emph{Phys. Rev. B}, vol.~14, pp. 4854--4873, 1976.

\bibitem{Kurakado}
M.~Kurakado, ``{Possibility of high resolution detectors using superconducting
  tunnel junctions},'' \emph{Nucl. Instrumen. Methods}, vol. 196, pp. 275--277,
  1982.

\bibitem{Kaplan_loss}
S.~B. Kaplan, ``Acoustic matching of superconducting films to substrates,''
  \emph{J. Low Temp. Phys.}, vol.~37, pp. 343--365, 1979.

\end{thebibliography}

\bibliographystyle{IEEEtran}

\end{document}